\documentclass[aps,prl,showpacs,twocolumn,groupedaddress]{revtex4}
\usepackage[latin1]{inputenc}
\usepackage{graphicx,epsfig,epstopdf}

\begin{document}

\title{Quantum Statistics of Interacting Dimer Spin Systems}

\author{
Ch. R\"uegg,$^{1,2}$
B. Normand,$^{3}$
M. Matsumoto,$^{4}$
Ch. Niedermayer,$^{2}$
A. Furrer,$^{2}$
K. W. Kr\"amer,$^5$
H.--U. G\"udel,$^5$
Ph. Bourges,$^6$
Y. Sidis,$^6$
and H. Mutka$^7$}

\affiliation{
$^1$ London Centre for Nanotechnology; Department of Physics and Astronomy; 
University College London; London WC1E 6BT; UK \\
$^2$ Laboratory for Neutron Scattering; ETH Zurich and Paul Scherrer 
Institute; 5232 Villigen PSI; Switzerland \\
$^3$ D\'epartement de Physique; Universit\'e de Fribourg; 1700 Fribourg; 
Switzerland \\
$^4$ Department of Physics; Faculty of Science; Shizuoka University; 
Shizuoka 422--8529; Japan \\
$^5$ Department of Chemistry and Biochemistry; University of Berne; 
3000 Bern 9; Switzerland \\
$^6$ Laboratoire L\'eon Brillouin (CEA--CNRS); CEA--Saclay; 91191 
Gif--sur--Yvette Cedex; France \\
$^7$ Institut Laue Langevin; BP 156; 38042 Grenoble Cedex 9; France \\ }

\date{\today}

\begin{abstract}
The compound TlCuCl$_3$ represents a model system of dimerized quantum 
spins with strong interdimer interactions. We investigate the triplet 
dispersion as a function of temperature by inelastic neutron scattering 
experiments on single crystals. By comparison with a number of theoretical 
approaches we demonstrate that the description of Troyer, Tsunetsugu, and 
W\"urtz [Phys. Rev. B {\bf 50}, 13515 (1994)] provides an appropriate 
quantum statistical model for dimer spin systems at finite temperatures, 
where many--body correlations become particularly important.
\end{abstract}


\maketitle

The temperature is without doubt the key control parameter in solid--state 
physics, for both historical and technical reasons. Historically, the study 
of thermal fluctuations created the discipline of statistical mechanics. 
As one technical example, the liquefaction of helium in 1908 enabled the 
discovery of superconductivity in 1911 \cite{Onnes11}. While control 
parameters such as the magnetic field or applied pressure may seem 
rather abstract by comparison, in certain classes of system they have a 
profound effect on the quantum mechanical fluctuations of the dominant 
degrees of freedom. In this respect, dimer spin systems such as 
BaCuSi$_2$O$_6$ \cite{Sasago97, Jaime04, Sebastian05}, TlCuCl$_3$ 
\cite{Cavadini01, Tanaka01, Rueegg03, Oosawa03, Rueegg04}, and 
Cs$_3$Cr$_2$Br$_9$ \cite{Grenier04} have recently attracted considerable 
attention as they exhibit both field-- and (in TlCuCl$_3$) pressure--induced 
magnetic quantum phase transitions. The ordered phases have also been 
described as a novel state resulting from Bose--Einstein condensation 
of magnons \cite{Nikuni00, Rice02}. 

To date the primary experimental and theoretical studies of these systems 
have focused on elucidating the parameter--dependence of the ground and 
excited states at ``$T = 0$~K'' \cite{Jaime04, Sebastian05, Tanaka01, 
Rueegg03, Oosawa03, Rueegg04, Grenier04}. However, the temperature--dependence 
of the spin--spin correlation function $S($\boldmath$Q$\unboldmath$,\omega)$ 
in such dimer--based compounds is also of fundamental interest. At finite 
temperatures, thermal fluctuations populating the excited triplet states 
above the energy gap, $\Delta \approx 7.5$~K, in the spin excitation 
spectrum of TlCuCl$_3$ become significant and compete with the intrinsic 
quantum fluctuations. The absence of a classical ordering transition 
in zero field and at ambient pressure, and the strong interdimer interactions, 
make TlCuCl$_3$ an ideal candidate for an experimental investigation of 
the quantum statistical description for such a spin system.

In TlCuCl$_3$, the antiferromagnetic (AF) dimer unit is formed by the two 
$S = 1/2$ magnetic moments in a pair of Cu$^{2+}$ ions, and has a singlet 
ground state (total spin $S = 0$) with triplet ($S = 1$) excited states which 
may be considered as bosonic quasiparticles. Only in the high--temperature 
limit, relative to the intradimer exchange constant $J \approx 56$~K 
\cite{Cavadini01, Matsumoto02}, is the system better described in terms 
of $S = 1/2$ fermions whose interactions are weak compared to the thermal 
fluctuations, and AF dimer correlations are strongly suppressed. In the 
intermediate regime, $\Delta < T < J$, a singlet--triplet description 
remains valid but the number of excited triplets becomes significant. Here 
the effect of interparticle interactions, particularly the intrinsic 
hard--core repulsion on each dimer bond, defines a challenging experimental 
and theoretical scenario. Temperature--dependent properties of dimer spin 
systems have to date been measured only in compounds of weakly interacting 
dimers with a spin gap large compared to the triplet bandwidth, including 
BaCuSi$_2$O$_6$ \cite{Sasago97}, KCuCl$_3$ \cite{Cavadini00}, and 
Cu(NO$_3$)$_2$$\cdot$2.5D$_2$O \cite{Xu00}, where many--body correlation 
effects are less pronounced. The description of these data was restricted 
to RPA theory or phenomenological modeling.

In this Letter we present the results of a recent inelastic neutron 
scattering (INS) study of the spin dynamics in TlCuCl$_{3}$. The primary 
results include a comprehensive experimental survey of the triplet excitations 
up to temperatures approaching the intradimer interaction strength, 
$\Delta \ll T_{\rm max} \approx 0.7 J$, and their theoretical modeling by 
quantum statistical considerations applied to the known zero--temperature 
states (i.e.~with no additional fitting parameters). We characterize the 
evolution of the spin gap, $\Delta(T)$, which is of particular relevance 
to the quantum critical behavior of the compound, and shows a pronounced 
increase with temperature. We demonstrate that the data support strongly 
the Troyer--Tsunetsugu--W\"urtz (TTW) Ansatz for the quantum statistics of 
general dimer spin systems \cite{Troyer94}. Our results may further be 
compared with those obtained for weakly interacting dimer compounds (above) 
and are closely related to extensive investigations of another class of 
gapped quantum magnets, $S = 1$ Haldane chains, which have been studied in 
model compounds such as CsNiCl$_{3}$ \cite{Kenzelmann02}.

Thermal renormalization of the dispersion, spectral weight and damping of 
the triplet modes in KCuCl$_3$, which shares the crystal structure and 
exchange coupling scheme of TlCuCl$_3$, are adequately described within 
a self--consistent RPA theory including higher--order corrections 
\cite{Cavadini00}. The considerably stronger interdimer interactions 
make this approach less successful for TlCuCl$_3$. We have therefore 
applied an extension of the bond--operator description of the compound 
\cite{Matsumoto02} to finite temperatures, a procedure requiring 
\cite{rnr} the inclusion of thermal population factors $n(E_{\alpha}
($\boldmath$q$\unboldmath$),\beta)$ in the mean--field (MF) equations
defining the singlet density $\bar{s}$ and the triplet chemical potential 
$\mu$; here $E$ denotes the magnon energy, \boldmath$q$\unboldmath~its
wave vector, $\beta = (k_{\rm B}T)^{-1}$, and $\alpha = -1,0,+1$ represents 
the $S_z = -1,0,+1$ triplet modes. We assume explicitly that there are 
no additional factors which may change with temperature, such as 
renormalization of the intra-- and interdimer interaction parameters 
due to phonons \cite{Uhrig98}. 

\begin{figure}[t!]
\includegraphics[width=0.35\textwidth]{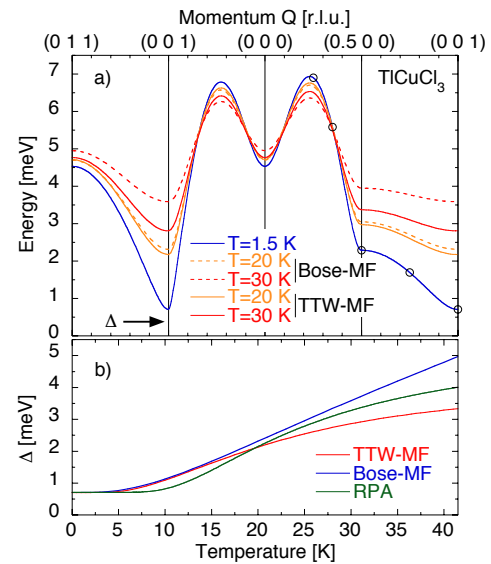}
\caption{\small (a) Finite--$T$ triplet dispersion for TlCuCl$_{3}$ 
from statistical models described in the text. $\Delta$ denotes the 
minimum singlet--triplet gap, \boldmath$Q$\unboldmath$=$\boldmath$q 
$\unboldmath$+$\boldmath$\tau$\unboldmath~the momentum, where 
\boldmath$\tau$\unboldmath~is a reciprocal--lattice vector, and open circles 
the \boldmath$Q$\unboldmath~values presented in Fig.~\ref{Tdep_disp}. 
(b) Temperature--dependence of $\Delta(T)$ from TTW--MF, Bose--MF, and 
RPA approaches. }
\label{Disptheory}
\end{figure}

The singlet state and each of the triplets on a single dimer obey 
an exclusion constraint and are described appropriately as hard--core 
bosons. Their statistical properties have no straightforward extension to 
reciprocal space and no exact expression exists for a triplet thermal 
occupation function. However, the effects of the constraint were 
introduced by Troyer and coworkers \cite{Troyer94} in the form of a 
statistical reweighting of the number of free--boson states of the 
$N$--dimer system to account for their complete local exclusion. The 
reweighting constitutes a suppression of state availability in all 
magnon occupation sectors $M$, being most significant at large $M$, 
and returns the effective single--dimer free energy 
\begin{equation}
\tilde f = - \frac{1}{\beta} \ln \left[ 1 + \sum_{\alpha} z_{\alpha}(\beta)
\right] 
\label{efe}
\end{equation}
with $z_{\alpha}(\beta) = \frac{1}{N} \sum_{\mbox{\boldmath$q$\unboldmath}} 
e^{-\beta E_{\alpha} (\mbox{\boldmath$q$\unboldmath})}$ the partition 
function for triplet mode $\alpha$. The effective occupation function for 
this statistical Ansatz (denoted TTW--MF) is then 
\begin{equation}
n(E_{\alpha}({\mbox{\boldmath$q$\unboldmath}}), \beta) = \frac{e^{-\beta 
E_{\alpha}({\mbox{\boldmath$q$\unboldmath}})}}{1 + \sum_{\alpha} z_{\alpha}
(\beta)}.
\label{TTW_n}
\end{equation}
Equation (2) reproduces both the low-- and high--temperature limits expected 
for the constrained boson system \cite{Troyer94}, namely $n \sim e^{-\beta 
E_{\alpha} (\mbox{\boldmath$q$\unboldmath})}$ at low $T$ (bosonic, $M$ small) 
and $n \sim 1/4$ at high $T$ (equipartition, $M$ large), and we will test 
below the accuracy of exclusion implementation by statistical reweighting 
for the regime between these limits. For comparison we will show also the 
results (denoted Bose--MF) obtained by enforcing the constraint at the 
Holstein--Primakoff level \cite{Matsumoto02} and considering a conventional 
bosonic magnon occupation factor. Fig.~\ref{Disptheory}(a) shows theoretical 
predictions for the thermal renormalization of the triplet dispersion: both 
approaches display a narrowing of the magnon band which arises because 
interdimer triplet hopping is obstructed by the exclusion condition when the 
concentration ($M/N$) of thermally excited triplets becomes significant.

The gap $\Delta$ at the dispersion minimum, which becomes 
the AF zone center in the magnetically ordered high--field and --pressure 
phases, deserves special attention. Fig.~\ref{Disptheory}(b) shows its 
thermal evolution as obtained from the above models, 
and from the RPA treatment applied to KCuCl$_3$ in Ref.~\cite{Cavadini00}. 
The Bose--MF approach shows a very strong renormalization at higher 
temperatures due to the possibility of unlimited boson occupation. This 
effect is considerably weaker in the TTW--MF calculations, where it is 
limited by the exclusion encoded in the statistics. These approaches 
coincide at low temperatures, where the triplet occupation is too dilute 
for their hard--core nature to be relevant. By contrast, in the RPA 
treatment the onset of a perceptible bandwidth narrowing requires much 
higher temperatures. 

\begin{figure}[t!]
\begin{minipage}[t!]{0.5\textwidth}
\hspace{-1.45cm}
\includegraphics[width=0.673\textwidth]{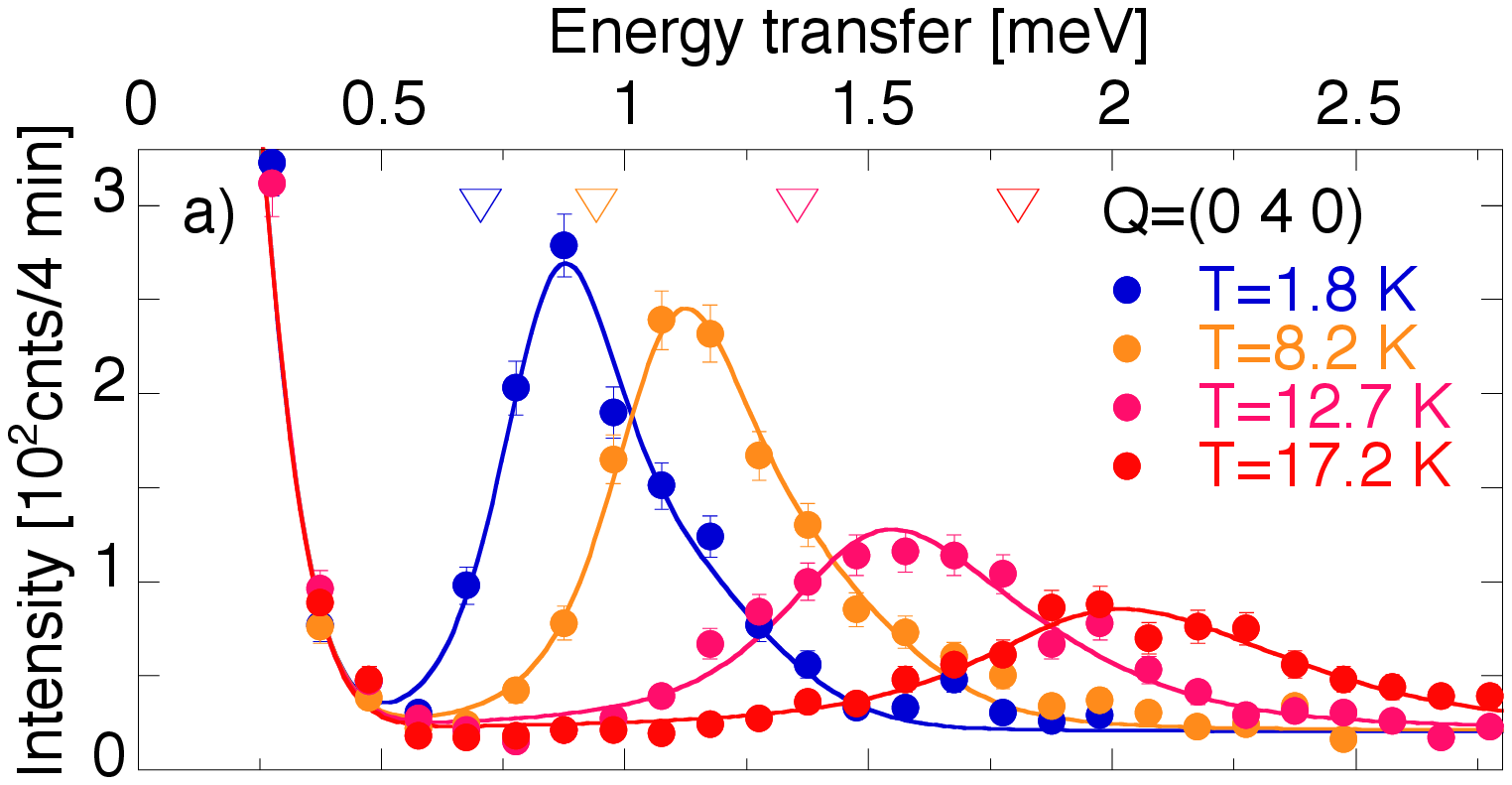}
\end{minipage}
\begin{minipage}[t!]{0.5\textwidth}
\includegraphics[width=0.85\textwidth]{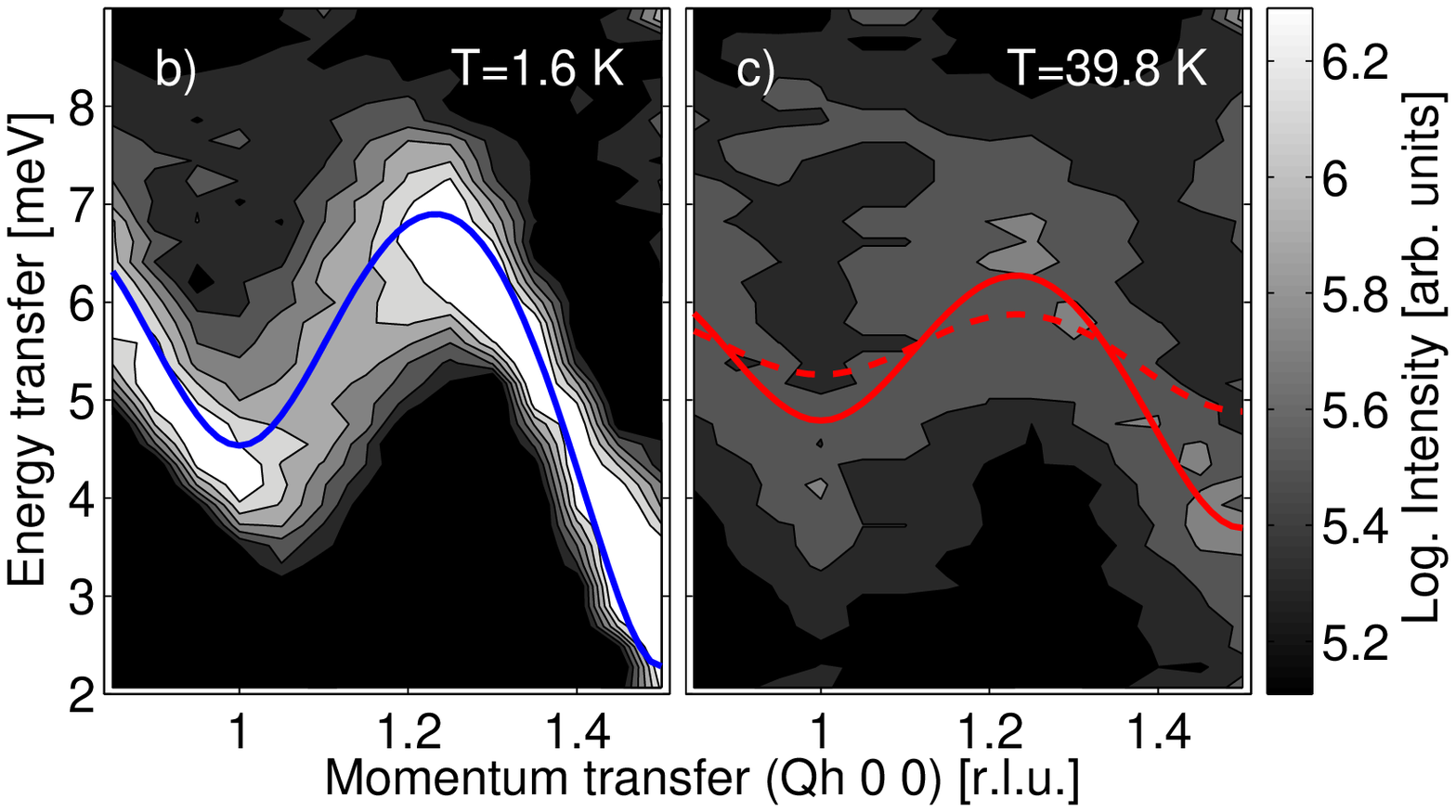}
\end{minipage}
\caption{\small (a) INS spectra for TlCuCl$_3$ measured on IN14 at 
\boldmath$Q$\unboldmath$=$(0~4~0) r.l.u. Solid lines are Lorentzian--based 
least--squares fits including full 4D--resolution convolution (peak 
asymmetry) and triangles indicate the physical excitation energy. (b) and (c) 
INS intensity contour plots measured at 1.6 K and 39.8 K, compared with 
theoretical curves by the Bose--MF (dashed) and TTW--MF (solid lines) 
descriptions.}
\label{TTW_BOSE_RPA}
\end{figure}

These theoretical results clearly motivate INS measurements of the triplet 
dispersion relation as a function of temperature at zero and finite field.
We have conducted INS experiments on single crystals of TlCuCl$_3$ at $H = 
0$. These were performed on the triple--axis spectrometers 
TASP (SINQ), IN14 (ILL), and 2T (LLB), working in constant--$k_{f}$ mode 
with a focusing PG analyzer/monochromator. While 2T is operated with thermal 
neutrons of $k_f = 2.662$ \AA$^{-1}$ (14.7 meV) and $1.970$ \AA$^{-1}$ (8.0 
meV), TASP and IN14 use moderated (cold) neutrons with a characteristic 
spectrum shifted to lower energies, $k_f = 1.506$ \AA$^{-1}$ (4.7 meV). A 
cooled beryllium or PG filter is positioned between the analyzer and the 
sample, contained in a standard cryostat providing $T \ge 1.5$ K. The 
(horizontal) collimation adopted on the cold instruments was 60'--open--open 
and 40'--open--open, yielding an energy resolution of $0.2$ meV (full width 
at half maximum height) in both cases. Corresponding values for the thermal 
instrument without additional collimation were 1.2 meV and 0.5 meV. 

\begin{figure}[t!]
\includegraphics[width=0.35\textwidth]{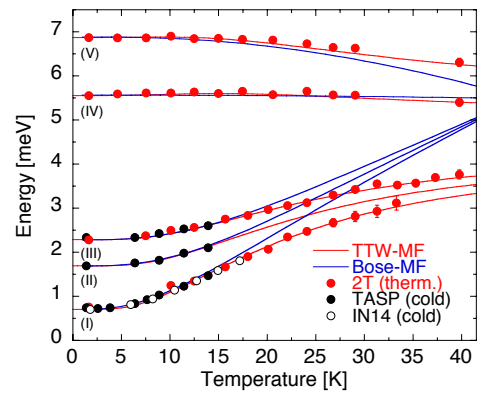}
\caption{\small Temperature--dependence of the excitation energies at (I) 
\boldmath$Q$\unboldmath$=$(0~0~1) (closed black circles and closed red 
circles) and \boldmath$Q$\unboldmath$=$(0~4~0) (open black circles), (II) 
\boldmath$Q$\unboldmath$=$(0.25~0~-1.5), (III) \boldmath$Q$\unboldmath$ 
=$(0.5~0~-2), (IV) \boldmath$Q$\unboldmath$=$(1.35~0~0), and (V) 
\boldmath$Q$\unboldmath$=$(1.25~0~0) (all in r.l.u.). Data from different 
instruments as indicated. Solid lines are obtained from TTW--MF and 
Bose--MF calculations.}
\label{Tdep_disp}
\end{figure}

Typical INS spectra for the triplet modes in TlCuCl$_3$ measured for a number 
of temperatures at the band minimum (\boldmath$Q$\unboldmath$\,=\,$(0~4~0) 
r.l.u.) are presented in Fig.~\ref{TTW_BOSE_RPA}(a). Upward renormalization
of the excitation energy, the onset of damping, and reduction in the 
integrated intensity are observed. Figs.~\ref{TTW_BOSE_RPA}(b) and (c) 
contrast the INS intensities measured at 1.6 and 39.8 K along ($Q_{h}$~0~0). 
The complete 
experimental survey, including representative \boldmath$Q$\unboldmath~values 
spanning the entire bandwidth, is presented in Fig.~\ref{Tdep_disp} 
and compared with the model calculations. For instrumental reasons the spin 
gap $\Delta$ must be measured in several experimental configurations to 
cover the full temperature range. The point \boldmath$Q$\unboldmath$
\,=\,$(1.35~0~0) r.l.u. is of particular interest: here the excitation 
energy corresponds approximately to $J$, and indeed it is nearly 
temperature--independent. The downward renormalization of the dispersion 
maximum is evident. The continuity between data sets from different 
instruments and \boldmath$Q$\unboldmath~values underlines the accuracy of 
the experimental procedure, including the data analysis (below).

With regard to the calculated fits, the TTW--MF model is clearly favored 
by the experimental observations in the range $20$ K $\le T \le 40$ K 
(Figs.~\ref{TTW_BOSE_RPA}(c) and \ref{Tdep_disp}) where the thermal 
renormalization changes less rapidly due to the constrained magnon 
population. The Bose--MF approach fails markedly in this respect. For 
clarity the RPA predictions are not shown in Fig.~\ref{Tdep_disp}, and 
we state only that these fail in particular to reproduce the low--temperature 
behavior [cf.~Fig.~\ref{Disptheory}(b)]. The agreement between the TTW--MF 
result and the experimental data is excellent over the full temperature range 
of the investigation. We stress that no free parameters are introduced with 
respect to the low--temperature dispersion \cite{Cavadini01, Matsumoto02}. 

We turn to a discussion of the excitation line width. For modes of infinite 
lifetime, $S($\boldmath$Q$\unboldmath$,\omega)$ is a $\delta$--function in 
energy ($E = \hbar \omega$), which within instrumental resolution is the 
case for the triplet modes measured in TlCuCl$_3$ at $T = 1.6$ K. However, 
at elevated temperatures a finite intrinsic line width (damping) is observed, 
and data analysis requires knowledge of the spectral function. Different 
$O(n)$ rotor models give a Lorentzian line shape in one dimension (1D) and 
at $T < \Delta$ \cite{Damle98}, but neither limit is appropriate here. By 
contrast, the damped harmonic oscillator (DHO) line shape used frequently 
for phonons \cite{Fak97} is identical to a double Lorentzian if the real 
excitation energy is defined as $E_{\rm DHO} ($\boldmath$q$\unboldmath$)^2 
= E($\boldmath$q$\unboldmath$)^2 + \Gamma^2$, and thus depends on the damping 
$\Gamma$ (half Lorentzian width at half maximum height). The DHO definition 
has been used to fit the finite--$T$ triplet excitations in Haldane chains 
\cite{Kenzelmann02}, and also for spin waves \cite{Forster75}. 

In the Bose--MF and TTW--MF approaches the energy is not renormalized by 
line shape effects, consistent with the Lorentzian assumption. While a line 
width $\Gamma$ may be computed from the phase space available for magnon decay 
and scattering, these considerations provide no proof of either a Lorentzian 
or a DHO form for triplet excitations in 3D, and nor is one available 
in the literature. We proceed on the basis of experiment: the INS measurements 
show that a Lorentzian yields an accurate description of the data. The spectra 
[Fig.~\ref{TTW_BOSE_RPA}(a)] were fitted with this line shape and the real 
excitation energy defined as $E($\boldmath$q$\unboldmath$)$. We suggest that 
this definition provides the correct description of the spectral function at 
finite temperatures in dimer spin systems. This conclusion is reinforced by 
the temperature--dependence of the dispersion maxima, which show a downward 
renormalization with increasing $T$. This behavior is consistent with the 
Lorentzian model, whereas $E_{\rm DHO} ($\boldmath$q$\unboldmath$)$ 
describes the maxima rising due to the strong increase of $\Gamma$ with 
temperature. Qualitatively, this result may be traced once again to the 
hard--core nature of the magnon excitations.

\begin{figure}[t!]
\includegraphics[width=0.35\textwidth]{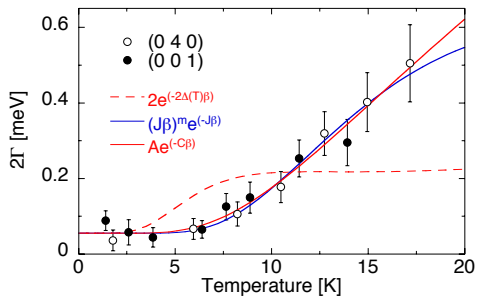}
\caption{\small Temperature--dependence of line width $2\Gamma$ for 
selected reciprocal--lattice points. Solid lines denote fits to 
the data as shown in the legend ($A$=3(1), $C$=2.7(5), and $m$=2.04(7)). The 
low--$T$ offset of 0.05 meV is attributed to uncertainty in the 
resolution deconvolution.}
\label{FWHM_T_fit}
\end{figure}

Fig.~\ref{FWHM_T_fit} shows the damping of the triplet modes, which 
is clearly activated. This behavior is expected from any model based 
on enhanced triplet--triplet scattering as the origin of a finite 
quasiparticle lifetime. We are unable to distinguish between fits  
using a variable activation energy or a variable power--law prefactor 
to an exponential with a constant activation energy of $J$ \cite{Xu00} 
(Fig.~\ref{FWHM_T_fit}). However, the data is definitely incompatible 
with an activation energy of $2 \Delta (T)$, as would be expected from 
magnon decay terms present at third order in the triplet hopping 
Hamiltonian, and these processes may be excluded. 

With the present investigation we would hope to initiate further studies. 
Detailed theoretical predictions for the finite--temperature spectral 
function of gapped excitations in 3D dimer spin systems are required. 
Experimental studies at magnetic fields around the critical field $H_{c}$ 
and at finite temperatures, i.e.~in the quantum critical regime, are 
further expected to reveal new physics.

In summary, we have investigated the thermal renormalization of triplet 
excitations in the strongly interacting dimer spin system TlCuCl$_3$ by 
means of INS. We find excellent agreement between the TTW--MF Ansatz and 
the complete data set, which was taken up to $T \approx 0.7 J$. The 
Bose--MF approach is in comparable agreement for $T < 20$ K and may 
therefore be considered as a suitable low--$T$ approximation. However, 
at higher temperatures, where many dimer triplets are excited, their 
hard--core nature becomes manifest and an accurate quantum statistical 
description requires the TTW Ansatz.

We thank T.M. Rice and M. Sigrist for many valuable discussions. We are 
further indebted to R. Cowley, B. Dorner, D. McMorrow, and B. Roessli for 
helpful comments. This work was performed in part at the Swiss spallation 
neutron source, SINQ, at the Paul Scherrer Institute, and was supported by 
the Swiss National Science Foundation and the NCCR MaNEP.

\end{document}